\documentclass[aps,prb,a4paper,notitlepage,noshowpacs,10pt]{revtex4-1}

\usepackage{graphicx}

\begin{document}

\title{AC Josephson effect between two superfluid time crystals}

\author{S.~Autti$^{1,2\ast}$}

\author{P.J. Heikkinen$^{1,3}$}

\author{J.T. M\"akinen$^{1,4,5}$}

\author{G.E.~Volovik$^{1,6}$}

\author{V.V. Zavjalov$^{1,2}$}

\author{V.B.~Eltsov$^{1 \dagger}$}

\affiliation{$^1$Low Temperature Laboratory, Department of Applied Physics, Aalto University, POB 15100, FI-00076 AALTO, Finland. $\dagger$vladimir.eltsov@aalto.fi \\
$^2$Department of Physics, Lancaster University, Lancaster LA1 4YB, UK. *s.autti@lancaster.ac.uk \\
$^3$Department of Physics, Royal Holloway, University of London, Egham, Surrey, TW20 0EX, UK. \\
$^4$Department of Physics, Yale University, New Haven, CT 06520, USA\\
$^5$Yale Quantum Institute, Yale University, New Haven, CT 06520, USA\\
$^6$L.D. Landau Institute for Theoretical Physics, Moscow, Russia}

\maketitle




\textbf{
Quantum time crystals are systems characterised by spontaneously emerging periodic order in the time domain \cite{PhysRevLett.109.160401}. A range of such phases has been reported (e.g. reviews \citenum{Sacha_2017,else2019discrete}). The concept has even been discussed in popular literature \cite{Popular1,Popular2,Ball_2018,Ball_2018_2}, and deservedly so: while the first speculation on a phase of broken time translation symmetry did not use the name ``time crystal'' \cite{Andreev1996}, it was later adopted from 1980's popular culture \cite{Ball_2018}. For the physics community, however, the ultimate qualification of a new concept is its ability to provide predictions and insight. Confirming that time crystals manifest the basic dynamics of quantum mechanics is a necessary step in that direction. We study two adjacent quantum time crystals experimentally. The time crystals, realised by two magnon condensates in superfluid $^3$He-B, exchange magnons leading to opposite-phase oscillations in their populations -
-- AC Josephson effect\cite{30002064863} --- while the defining periodic motion remains phase coherent throughout the experiment.}

\begin{figure}[htb!]
\centerline{\includegraphics[width=1\linewidth]{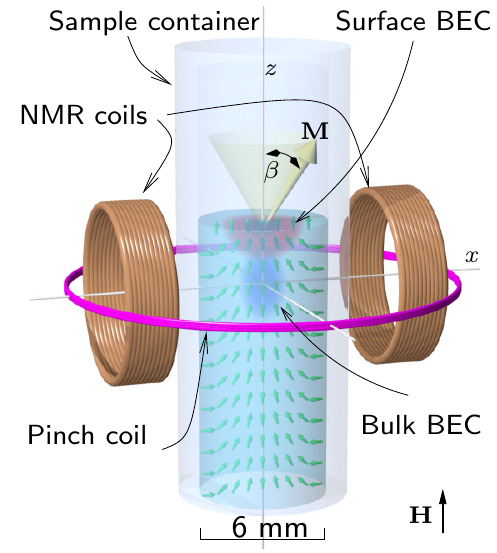}}
\caption{\textbf{Experimental setup:} Quartz-glass sample container cylinder is filled partially with superfluid $^3$He-B, leaving a free surface of the superfluid approximately 3~mm above the centre of the surrounding coil system. The space above the free surface is vacuum due to the vanishing vapour pressure of $^3$He at sub-mK temperatures. Magnons can be trapped in this configuration in two separate locations, in bulk (coloured blue) and touching the free surface (coloured red). Transverse NMR coils are used both for RF pumping of magnons into the BECs, and for recording the induced signal from the coherently precessing magnetisation $\mathbf{M}$. The amplitude of the recorded signal is proportional to $\beta$, the tipping angle of $\mathbf{M}$, and its frequency corresponds to the precession frequency of the condensate. The condensates are trapped by the combined effect of the distribution of orbital anisotropy axis of the superfluid (green arrows) via spin-orbit coupling, and a minimum of the external 
magnetic field created using a pinch coil. 
The external field $\mathbf{H}$ is oriented along the $z$ axis of the sample container.
}
\label{cell}
\end{figure}

A magnon Bose-Einstein condensate (BEC) in superfluid $^3$He-B is a macroscopic quantum state described by a simple wavefunction $\Psi = |\Psi| e^{- i (\mu t/\hbar +\varphi_\mathbf{M})}$, where $\mu$ is the chemical potential \cite{magnon_BEC_review}. Magnons are quanta of transverse spin waves, corresponding to magnetisation $\mathbf{M}$ that precesses at frequency $f=\mu/(2 \pi \hbar)$ around the external magnetic field $\mathbf{H}$, starting from initial phase $\varphi_\mathbf{M}$. Each magnon carries $-\hbar$ of spin, yielding total number of magnons $N\propto|\Psi|^2\propto \beta^2$, where $\beta$ is the deflection angle of $\mathbf{M}$ from the equilibrium direction along $\mathbf{H}$. Here we assumed that $\beta$ is small, which is satisfied in all the experiments presented in this Letter. In general, Bose-Einstein condensates are an established platform for studying both DC \cite{HPD_Josephson} and AC Josephson effects \cite{Levy2007579,Valtolina1505,Abbarchi2013275}.

The manifest feature of quasiparticle condensation is the emergence of spontaneous coherence of the precession frequency and phase \cite{1992_ppd,2000_ppd,1984_hpd_exp,Bunkov_2010}. In our system it also guarantees that the precession period forms uncontrived: for example, it is possible to pump magnons to a higher-frequency level in a confining trap, from which magnons then fall to the ground level, thereby choosing a period independent of that of the drive\cite{PhysRevLett.120.215301}. This periodic, observable motion constitutes the essence of a time crystal. It is detected in our system using nuclear magnetic resonance techniques (NMR), based on coupling the precessing magnetisation to nearby pick-up coils (see Fig.~\ref{cell}).

In NMR experiments, magnon time crystals in superfluid $^3$He-B are characterised by two timescales \cite{Volovik2013}. The first timescale $\tau_E \sim 0.1~$s describes how quickly precession in the condensate becomes coherent, following the pumping of incoherent magnons. The second timescale $\tau_N$ is magnon lifetime. The system reaches exact particle conservation in the limit $\tau_N \rightarrow \infty$, which in an isolated sample container is approached exponentially as temperature decreases. In practice there are losses in the pick-up coils that are coupled to the precessing spins in order to control and observe the condensate\cite{magnon_relax}. It is also impractical to carry out experiments at the lowest achievable temperature \cite{2000_ppd}. It is therefore necessary to allow for a finite $\tau_N$, which is acceptable as long as $\tau_N \gg \tau_E$. 

One can then either observe the phenomena that emerge during this slow decay, or compensate for the losses by pumping the system continuously. Under continuous pumping, the condensate spontaneously finds a magnon number that corresponds to the chemical potential $\mu$ set by the pumping frequency \cite{volovik_bunkov_qball,magnon_trap_mod}. On the other hand, free decay with the pumping turned off is particularly instrumental for studying dynamics and interactions of magnon-BEC time crystals as it removes the need to distinguish potential artefacts of the external driving force. These features make magnon condensates an ideal laboratory system for studying time crystals, their dynamics, and related emergent phenomena.

\begin{figure*}
\centerline{\includegraphics[width=1\linewidth]{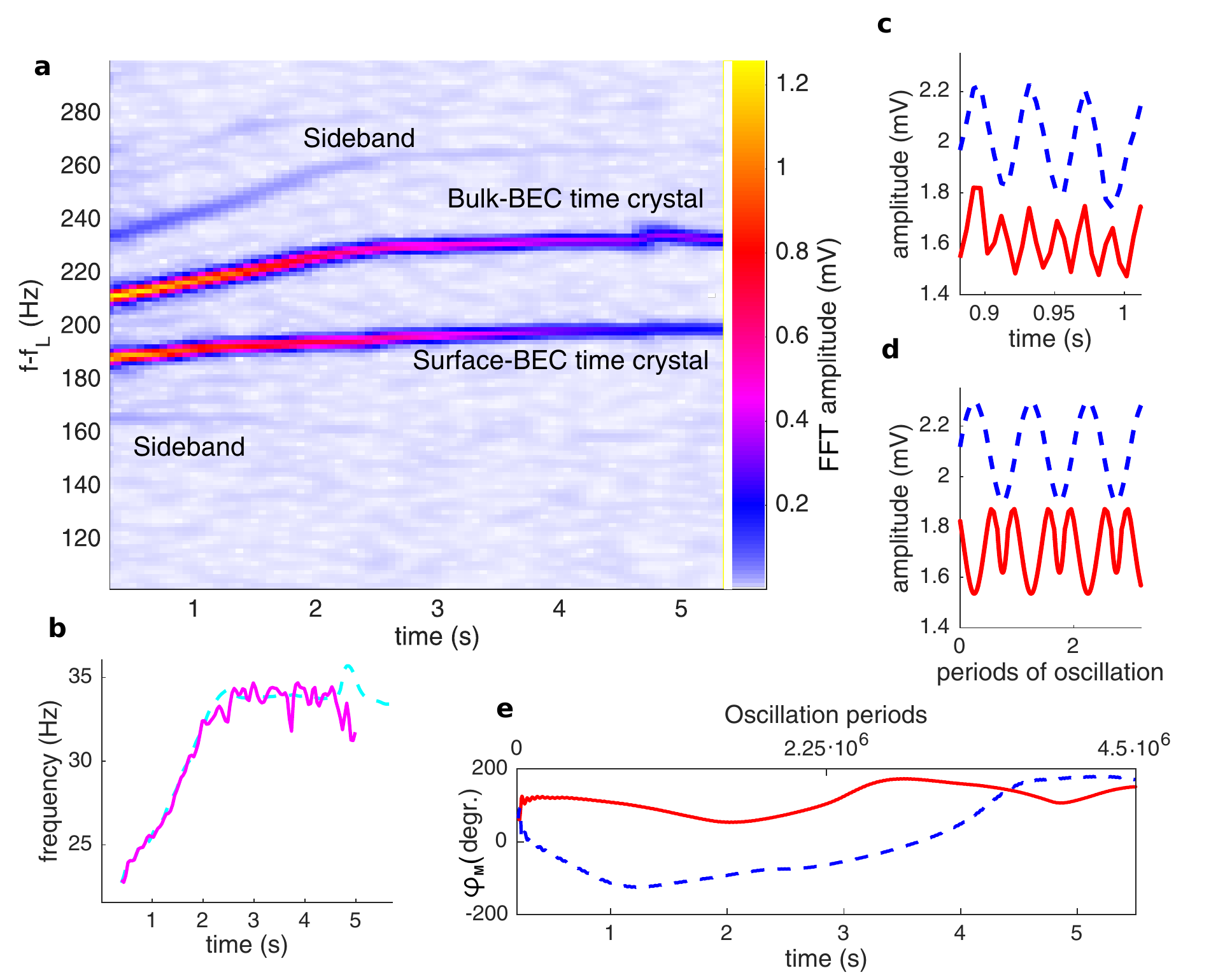}}
\caption{\textbf{Time crystal AC Josephson effect:} ({\bf a}) Two co-existing magnon-BEC time crystals, created with an RF drive pulse at $t=0$, are seen as peaks in the Fourier spectrum of the signal recorded from the NMR coils. For clarity, the exciting pulse is left just outside the time window shown here. The FFT amplitude refers to the voltage measured from the pick-up coils after pre-amplification, and $f_L=833~$kHz is the Larmor frequency. The upper trace corresponds to the magnon-BEC time crystal in the bulk, and the lower trace to the time crystal touching the free surface. The bulk trap is the more flexible of the two, and the bulk time crystal frequency hence increases during the decay more than that of the surface crystal. Population oscillations between the time crystals result in amplitude oscillations of the two signals, seen as two side bands. ({\bf b}) The changing frequency difference of the two time crystals (cyan dash line) matches the frequency of the population oscillations between them,
 extracted from the bulk crystal side band (magenta line). ({\bf c}) Direct fits to the recorded signal at the frequency of the bulk crystal (blue dash line) and the surface crystal (solid red line) reveal AC Josephson oscillations of population between the two crystals: the opposite-phase component of the amplitude oscillation is attributed to the AC Josephson effect, while the in-phase component in the surface crystal signal is due to trapping potential changes imposed by the bulk crystal oscillations. ({\bf d}) The quasi-static numerical simulation reveals that changes in the bulk crystal population (blue dash line) distort the trapping potential, adding an additional component to the calculated signal from the surface crystal (solid red line). ({\bf e}) The azimuthal angle of precessing magnetisation $\varphi_\mathbf{M}$ in the rotating frame in the bulk crystal (blue dash line) and the surface crystal (solid red line) are extracted by feeding the raw signal to a software lock-in amplifier, locked to the 
corresponding frequency traces in panel {\bf a}.  
}
\label{Josephson_effect1}
\end{figure*}

Magnons are trapped in the middle of the sample container cylinder by the combined effect of the superfluid order parameter distribution (``texture''), and an axial minimum in the magnetic field\cite{thuneberg_texture,magnon_relax}. This yields an approximately harmonic three-dimensional trap. Introducing a free surface, located 3~mm above the field minimum, modifies the textural trap creating two local minima and, hence, splits the magnon BEC spectrum  into two physical locations: (1) The bulk trap remains approximately harmonic. (2) An additional surface trap emerges, see Fig.~\ref{cell}. States in the measured experimental spectrum can be identified based on their dependence on the profile of the magnetic trap, controlled by changing the current in the pinch coil. In what follows we concentrate on studying the lowest-energy state of each of the two traps.


The measurement begins by populating both the bulk and the surface traps using a radio-frequency (RF) pulse. The duration, frequency, and amplitude of the pulse are chosen so that the two traps are populated approximately equally. The signal recorded from the pick-up coils is then visualised using time-windowed Fourier analysis (Fig.~\ref{Josephson_effect1}a). When the pumping is turned off, the condensate populations decrease slowly due to dissipation as seen in the decrease of recorded signal amplitude from both condensates. The bulk condensate frequency is also increasing during the decay by $20~$Hz. This is because the textural trap feels the local pressure of magnons via spin-orbit interaction, and thereby becomes expanded when the number of magnons is large, making the trap shallower \cite{magnon_trap_mod,PhysRevB.97.014518,volovik_bunkov_qball}. The surface trap is rigidified by the textural boundary condition set by the free surface. Hence, the frequency of the surface condensate only changes by 7~Hz.
 The period (frequency) of each 
condensate is independent of the drive pulse features during the decay. This justifies calling the observed states time crystals. In what follows we refer to the magnon-condensate time crystals simply as ``(time) crystals''.

In addition to the traces corresponding to the time crystals, Fig~\ref{Josephson_effect1} also features two side bands. The side bands are separated from the main traces by the frequency difference between the two crystals, which changes slowly in time. We interpret the side bands as follows: The phase difference between the time crystals follows $\mathrm{d}(\varphi_\mathrm{bulk}-\varphi_\mathrm{surf})/\mathrm{d}t=-(\mu_\mathrm{bulk}-\mu_\mathrm{surf})/\hbar$, where the phase $ \varphi=-\mu t /\hbar + \varphi_\mathbf{M}$. The changing phase difference therefore drives an alternating Josephson supercurrent of constituent particles between the time crystals.
 This is seen as amplitude oscillations of the two signals, producing the side bands. The frequency of the oscillations is set by the difference of the time crystal precession frequencies, equal to the difference of their chemical potentials (Fig.~\ref{Josephson_effect1}b). These observations are characteristic to AC Josephson effect \cite{Levy2007579,Valtolina1505}. The remarkable advantage of the time crystal compared to superfluids and superconductors is that all four variables in the canonical Josephson equation ($\varphi_\mathrm{bulk}$,$\varphi_\mathrm{surf}$, $\mu_\mathrm{bulk}$,$\mu_\mathrm{surf}$) are now measured directly in the same experiment. 

We fit the measured signal, in short time windows, directly with two sine curves. This allows extracting the signal amplitude from each time crystal separately (Fig.~\ref{Josephson_effect1}c). The result shows that the bulk crystal signal amplitude (population) oscillates at a frequency equal to the frequency difference of the bulk and the surface time crystals. The surface crystal signal shows similar oscillations with the opposite phase, as expected for AC Josephson effect. The surface crystal however also features in-phase oscillations. Below we demonstrate, using numerical simulations, that the bulk crystal oscillation modifies the textural trapping potential around it periodically, and the modification propagates along the texture to change the shape of the surface time crystal. This shape change modifies the effective filling factor of the surface crystal between the pick-up coils, hence changing the resulting signal, while the magnon number in the surface time crystal is not affected by this 
modification.

It is worth emphasising that in the frame rotating with frequency $\mu/\hbar$, the (initial) azimuthal angle of magnetisation $\varphi_\mathbf{M}$ in each of the two time crystals remains stable over more than $10^6$ periods of oscillation (Fig.~\ref{Josephson_effect1}e), despite the population exchange, and the slow decay of overall magnon number. The azimuthal angle can be extracted by feeding the signal recorded from the pick-up coils to a software lock-in amplifier, locked to the frequency extracted from FFT analysis in Fig.~\ref{Josephson_effect1}a. The Josephson oscillations are filtered out by choosing a lock-in time constant longer than the Josephson frequency. The remaining drift is attributed to inaccuracy of the used reference frequency. The phase stability culminates the robustness of the time crystal, well-defined periodicity being the defining feature of broken time translation symmetry. 

Let us confirm that the trapping potential connects the two time crystals indirectly by building a self-consistent numerical simulation of the two underlying magnon condensates in the flexible trap (see Methods). The calculation qualitatively reproduces the remarkable features seen in the experiment (Fig.~\ref{Josephson_effect1}d): The signal from the surface time crystal shows twice shorter period than the signal from the bulk crystal. This is caused by changes in the shape of the surface trap, imposed by oscillations in the bulk crystal population. The calculation is quasi-static, meaning that the trap is assumed to adjust to changes in the magnon distribution instantaneously. This  explains why the signal from the surface crystal is aligned differently with the bulk crystal oscillations than observed in the experiment. Comparing the amplitude of the simulated oscillations with the experiment also supports the view that the observed Josephson oscillations in the two recorded experimental signals correspond 
to equal and opposite 
changes in the populations of the two time crystals. The oscillations of the trapping potential also directly change the precession frequencies of both condensates \cite{magnon_trap_mod,PhysRevB.97.014518}, thus changing the frequency difference in phase with the population changes. That should result in distortion of the sinusoidal population exchange, yielding additional side bands in Fig.~\ref{Josephson_effect1}. In practice this effect is too weak to be distinguished in the experiment.


In conclusion, we report an experimental realisation of two adjacent quantum time crystals that exchange constituent particles via the AC Josephson effect. The time crystals are created in a flexible trap in superfluid $^3$He-B, emerging as two spatially separate magnon BECs associated with coherent spin precession. The configuration of two interacting condensates is stabilised in the proximity of a free surface of the superfluid. The Josephson oscillations in the number of particles each time crystal contains are seen as opposite-phase amplitude variations in the measured signal. Flexibility of the trapping potential connects the two time crystals also indirectly providing additional interaction that results in an in-phase component of oscillation, as verified by numerical simulations. In the rotating frame, the azimuthal angle of each of the two time crystals remains stable and well defined in the course of all these perturbations. Notably, all the observables that characterise the AC Josephson effect, the 
time crystal precession phases and their chemical potentials, are directly measured in the same experiment. This relatively novel phase of matter therefore deserves its place in physicists' vocabulary. 

It remains an interesting task for future to study more sophisticated time crystal interactions. For instance, one could simulate the Hamiltonian of a Penrose-type ``gravitationally'' induced wave function collapse \cite{PhysRevLett.14.57} by allowing two time crystals in their flexible traps to collide. On the other hand, long-lived coherent quantum systems with tunable interactions, such as the robust time crystals studied here, provide a platform for building novel quantum devices based on spin-coherent phenomena\cite{Sato_2011}. For example, the dependence of the chemical potential on the time crystal populations, coupled by the Josephson junction, could be used as a shunting ``capacitor'' for the junction, forming an equivalent to the transmon qubit. Such devices based on macroscopic spin coherence could perhaps be implemented even at room temperature \cite{alex2019josephson}.

\section*{Acknowledgements}

This work has been supported by the European Research Council (ERC) under the European Union's Horizon 2020 research and innovation programme (Grant Agreement No. 694248). The experimental work was carried out in the Low Temperature Laboratory, which is a part of the OtaNano research infrastructure of Aalto University and of the European Microkelvin Platform. S. Autti acknowledges financial support from the Jenny and Antti Wihuri foundation, and P. J. Heikkinen that from the V\"{a}is\"{a}l\"{a} foundation of the Finnish Academy of Science and Letters.\newline
	
\section*{Competing interests}
The authors declare no competing interests.

\section*{Author contributions}
All authors contributed to writing the manuscript and discussing the results. Experiments were carried out and planned by S.A., J.T.M, P.J.H, V.V.Z, and V.B.E. Theoretical work was done by S.A., G.E.V, and V.B.E.

\section*{Methods}

We cool superfluid helium-3 down to $130~\mu$K ($0.14 ~T_\mathrm{c}$) using a nuclear demagnetisation cryostat\cite{Heikkinen2014}. Temperature is measured using quartz tuning forks~\cite{2007_forks, 2008_forks}. The superfluid transition temperature at 0~bar pressure is $T_\mathrm{c}=930~\mu$K. The superfluid is contained in a quartz-glass cylinder (radius 3~mm), placed in an external magnetic field of about 25~mT, aligned along the axis of the container. We emphasise that the results in the present Letter are not specific to this field or temperature: similar AC Josephson oscillations were observed down to 17mT and at temperatures up to $0.2~T_\mathrm{c}$. 

The sample container is  surrounded by transverse coils, needed for creating and observing the magnon condensate using NMR. First, the coils can be used to create a transverse radio-frequency (RF) field $H_\mathrm{rf}$, which tips magnetisation within the coils by a small amount. This allows pumping magnons into the sample. Second, the coils are used for recording the resulting coherent precession of magnetisation that induces an electromotive force (EMF) into the pick-up coils (Fig.~\ref{cell}). 


The free surface is located 3~mm above the centre of the magnetic field minimum. The distance of the free surface is determined by comparing the observed magnon spectrum with the numerical model described below, and confirmed by measuring the pressure of $^3$He gas in a calibrated volume that results from the removal of liquid from the originally fully-filled sample container. The free surface distorts the textural trap and splits the magnon spectrum into two physical locations, as detailed in the main text. Analysis of the whole observable spectrum will be published separately \cite{pjheikki_thesis}.

We simulate the magnon condensates in a quasi-static approximation using a two-step model following the lines of Refs.~\citenum{PhysRevB.97.014518,magnon_trap_mod,sautti_thesis}.  The first step is to calculate the trapping potential in the absence of magnons, and solve the corresponding magnon spectrum. This is achieved by minimising the free energy functional of the equilibrium superfluid \cite{thuneberg_texture,kopu_texture}, including the orienting effects of the magnetic field, sample container walls, and the free surface. The effect of the free surface is assumed to be described by the same parameter values that apply to solid walls. The magnetic field is calculated based on the known geometry of the coil system. In the absence of the free surface this results in an approximately harmonic trap for magnons \cite{magnon_relax,Heikkinen2014}. In the presence of the free surface the two spatially-separated components of the calculated spectrum semi-quantitatively correspond to those observed experimentally 
\cite{pjheikki_thesis}, but all the states touching the free surface are shifted upwards by roughly 150~Hz in the simulation as compared with the experimental spectra. While this means that the surface condensate shape is not described perfectly, it provides a more-than-sufficient starting point for the purposes of the present work. Finding detailed quantitative agreement remains a task for future studies. 

The second step in the model construction is to enable non-zero magnon density. The textural part of the trapping potential feels local magnon density due to spin-orbit interaction. For a fixed magnon distribution, this contribution can be included in the textural free energy minimisation. A self-consistent solution for given magnon number can then be found by fixed point iteration, as described in Refs.~\citenum{PhysRevB.97.014518,sautti_thesis}. Signal from the condensates is calculated according to the EMF induced in the pick-up coils due to the coherently precessing magnetisation in each condensate. We calibrate the simulation signal amplitude using the measured frequency shift as a function of signal amplitude \cite{PhysRevB.97.014518}. Josephson oscillations are 
emulated in our model by adding opposite-phase equal-amplitude oscillations of magnon number between the two condensates.

Magnons placed in the surface-touching condensate change the trap confining them less than those placed in the bulk condensate, as seen in Fig.~\ref{Josephson_effect1}a. This is because the boundary condition for the texture set by the free surface is orders of magnitude stronger than the effect of magnons. For simplicity, we therefore neglect the effect of the surface condensate population altogether in the calculation of the trapping potential. In our experiments the bulk condensate frequency is slightly higher than that of the surface-touching condensate. In the simulation we tune the current in the pinch coil such that the bulk condensate has the lowest frequency in the system, 50~Hz below the surface condensate frequency in the limit of zero magnons. This allows finding a self-consistent solution at all magnon numbers straightforwardly, as the self-consistency step in the simulation targets the the bulk condensate only. This simplification does not change the textural connection between the bulk 
condensate and the surface 
condensate, or the coupling of the condensates to the pick-up coils. 

\bibliography{TimeCrystalJ}

\begin{thebibliography}{10}
\expandafter\ifx\csname url\endcsname\relax
  \def\url#1{\texttt{#1}}\fi
\expandafter\ifx\csname urlprefix\endcsname\relax\def\urlprefix{URL }\fi
\providecommand{\bibinfo}[2]{#2}
\providecommand{\eprint}[2][]{\url{#2}}

\bibitem{PhysRevLett.109.160401}
\bibinfo{author}{Wilczek, F.}
\newblock \bibinfo{title}{Quantum time crystals}.
\newblock \emph{\bibinfo{journal}{Phys. Rev. Lett.}}
  \textbf{\bibinfo{volume}{109}}, \bibinfo{pages}{160401}
  (\bibinfo{year}{2012}).

\bibitem{Sacha_2017}
\bibinfo{author}{Sacha, K.} \& \bibinfo{author}{Zakrzewski, J.}
\newblock \bibinfo{title}{Time crystals: a review}.
\newblock \emph{\bibinfo{journal}{Reports on Progress in Physics}}
  \textbf{\bibinfo{volume}{81}}, \bibinfo{pages}{016401}
  (\bibinfo{year}{2017}).

\bibitem{else2019discrete}
\bibinfo{author}{Else, D.~V.}, \bibinfo{author}{Monroe, C.},
  \bibinfo{author}{Nayak, C.} \& \bibinfo{author}{Yao, N.~Y.}
\newblock \bibinfo{title}{Discrete time crystals}.
\newblock \emph{\bibinfo{journal}{arXiv:1905.13232}}  (\bibinfo{year}{2019}).

\bibitem{Popular1}
\bibinfo{author}{Wilczek, F.}
\newblock \bibinfo{title}{The exquisite precision of time crystals}.
\newblock \emph{\bibinfo{journal}{Scientific American}}
  \textbf{\bibinfo{volume}{November}} (\bibinfo{year}{2019}).

\bibitem{Popular2}
\bibinfo{author}{Gibney, E.}
\newblock \bibinfo{title}{The quest to crystallize time}.
\newblock \emph{\bibinfo{journal}{Nature}} \textbf{\bibinfo{volume}{543}},
  \bibinfo{pages}{164–166} (\bibinfo{year}{2017}).

\bibitem{Ball_2018}
\bibinfo{author}{Ball, P.}
\newblock \bibinfo{title}{In search of time crystals}.
\newblock \emph{\bibinfo{journal}{Physics World}}
  \textbf{\bibinfo{volume}{31}}, \bibinfo{pages}{29--33}
  (\bibinfo{year}{2018}).

\bibitem{Ball_2018_2}
\bibinfo{author}{Ball, P.}
\newblock \bibinfo{title}{Out of step with time}.
\newblock \emph{\bibinfo{journal}{Nature Materials}}
  \textbf{\bibinfo{volume}{17}}, \bibinfo{pages}{569} (\bibinfo{year}{2018}).

\bibitem{Andreev1996}
\bibinfo{author}{Andreev, A.~F.}
\newblock \bibinfo{title}{Bose condensation and spontaneous breaking of the
  uniformity of time}.
\newblock \emph{\bibinfo{journal}{Journal of Experimental and Theoretical
  Physics Letters}} \textbf{\bibinfo{volume}{63}}, \bibinfo{pages}{1018--1025}
  (\bibinfo{year}{1996}).

\bibitem{30002064863}
\bibinfo{author}{Jospehson, B.}
\newblock \bibinfo{title}{Possible new effect in superconducting tunneling}.
\newblock \emph{\bibinfo{journal}{Phys. Lett.}} \textbf{\bibinfo{volume}{1}},
  \bibinfo{pages}{251--253} (\bibinfo{year}{1962}).

\bibitem{magnon_BEC_review}
\bibinfo{author}{Bunkov, Y.~M.} \& \bibinfo{author}{Volovik, G.~E.}
\newblock \emph{\bibinfo{title}{Novel Superfluids}}, vol.~\bibinfo{volume}{1}
  (\bibinfo{publisher}{Oxford University Press, Oxford}, \bibinfo{year}{2013}).

\bibitem{HPD_Josephson}
\bibinfo{author}{Borovik-Romanov, A.~S.} \emph{et~al.}
\newblock \bibinfo{title}{Observation of a spin-current analog of the
  {Josephson} effect}.
\newblock \emph{\bibinfo{journal}{JETP Lett.}} \textbf{\bibinfo{volume}{47}},
  \bibinfo{pages}{1033--1037} (\bibinfo{year}{1988}).

\bibitem{Levy2007579}
\bibinfo{author}{Levy, S.}, \bibinfo{author}{Lahoud, E.},
  \bibinfo{author}{Shomroni, I.} \& \bibinfo{author}{Steinhauer, J.}
\newblock \bibinfo{title}{The a.c. and d.c. {Josephson} effects in a
  {Bose-Einstein} condensate}.
\newblock \emph{\bibinfo{journal}{Nature}} \textbf{\bibinfo{volume}{449}},
  \bibinfo{pages}{579--583} (\bibinfo{year}{2007}).

\bibitem{Valtolina1505}
\bibinfo{author}{Valtolina, G.} \emph{et~al.}
\newblock \bibinfo{title}{Josephson effect in fermionic superfluids across the
  {BEC-BCS} crossover}.
\newblock \emph{\bibinfo{journal}{Science}} \textbf{\bibinfo{volume}{350}},
  \bibinfo{pages}{1505--1508} (\bibinfo{year}{2015}).

\bibitem{Abbarchi2013275}
\bibinfo{author}{Abbarchi, M.} \emph{et~al.}
\newblock \bibinfo{title}{Macroscopic quantum self-trapping and {Josephson}
  oscillations of exciton polaritons}.
\newblock \emph{\bibinfo{journal}{Nature Physics}}
  \textbf{\bibinfo{volume}{9}}, \bibinfo{pages}{275--279}
  (\bibinfo{year}{2013}).

\bibitem{1992_ppd}
\bibinfo{author}{Bunkov, Y.~M.}, \bibinfo{author}{Fisher, S.~N.},
  \bibinfo{author}{Gu\'enault, A.~M.} \& \bibinfo{author}{Pickett, G.~R.}
\newblock \bibinfo{title}{Persistent spin precession in {$^{3}\text{He-B}$} in
  the regime of vanishing quasiparticle density}.
\newblock \emph{\bibinfo{journal}{Phys. Rev. Lett.}}
  \textbf{\bibinfo{volume}{69}}, \bibinfo{pages}{3092--3095}
  (\bibinfo{year}{1992}).

\bibitem{2000_ppd}
\bibinfo{author}{Fisher, S.} \emph{et~al.}
\newblock \bibinfo{title}{Thirty-minute coherence in free induction decay
  signals in superfluid {$^{3}\text{He-B}$}}.
\newblock \emph{\bibinfo{journal}{J. Low Temp. Phys.}}
  \textbf{\bibinfo{volume}{121}}, \bibinfo{pages}{303--308}
  (\bibinfo{year}{2000}).

\bibitem{1984_hpd_exp}
\bibinfo{author}{Borovik-Romanov, A.~S.}, \bibinfo{author}{Bun'kov, Y.~M.},
  \bibinfo{author}{Dmitriev, V.~V.} \& \bibinfo{author}{Mukharskii, Y.~M.}
\newblock \bibinfo{title}{{Long-lived induction signal in superfluid
  {$^{3}\text{He-B}$}}}.
\newblock \emph{\bibinfo{journal}{JETP Lett.}} \textbf{\bibinfo{volume}{40}},
  \bibinfo{pages}{1033--1037} (\bibinfo{year}{1984}).

\bibitem{Bunkov_2010}
\bibinfo{author}{Bunkov, Y.~M.} \& \bibinfo{author}{Volovik, G.~E.}
\newblock \bibinfo{title}{Magnon {Bose{\textendash}Einstein} condensation and
  spin superfluidity}.
\newblock \emph{\bibinfo{journal}{Journal of Physics: Condensed Matter}}
  \textbf{\bibinfo{volume}{22}}, \bibinfo{pages}{164210}
  (\bibinfo{year}{2010}).

\bibitem{PhysRevLett.120.215301}
\bibinfo{author}{Autti, S.}, \bibinfo{author}{Eltsov, V.~B.} \&
  \bibinfo{author}{Volovik, G.~E.}
\newblock \bibinfo{title}{Observation of a time quasicrystal and its transition
  to a superfluid time crystal}.
\newblock \emph{\bibinfo{journal}{Phys. Rev. Lett.}}
  \textbf{\bibinfo{volume}{120}}, \bibinfo{pages}{215301}
  (\bibinfo{year}{2018}).

\bibitem{Volovik2013}
\bibinfo{author}{Volovik, G.~E.}
\newblock \bibinfo{title}{On the broken time translation symmetry in
  macroscopic systems: Precessing states and off-diagonal long-range order}.
\newblock \emph{\bibinfo{journal}{JETP Letters}} \textbf{\bibinfo{volume}{98}},
  \bibinfo{pages}{491--495} (\bibinfo{year}{2013}).

\bibitem{magnon_relax}
\bibinfo{author}{Heikkinen, P.~J.} \emph{et~al.}
\newblock \bibinfo{title}{Relaxation of {$\text{Bose-Einstein}$} condensates of
  magnons in magneto-textural traps in superfluid {$^{3}\text{He-B}$}}.
\newblock \emph{\bibinfo{journal}{J. Low Temp. Phys.}}
  \textbf{\bibinfo{volume}{175}}, \bibinfo{pages}{3--16}
  (\bibinfo{year}{2014}).

\bibitem{volovik_bunkov_qball}
\bibinfo{author}{Bunkov, Y.~M.} \& \bibinfo{author}{Volovik, G.~E.}
\newblock \bibinfo{title}{Magnon condensation into a {Q-ball} in
  {$^{3}\text{He-B}$}}.
\newblock \emph{\bibinfo{journal}{Phys. Rev. Lett.}}
  \textbf{\bibinfo{volume}{98}}, \bibinfo{pages}{265302}
  (\bibinfo{year}{2007}).

\bibitem{magnon_trap_mod}
\bibinfo{author}{Autti, S.} \emph{et~al.}
\newblock \bibinfo{title}{Self-trapping of magnon {$\text{Bose-Einstein}$}
  condensates in the ground state and on excited levels: From harmonic to box
  confinement}.
\newblock \emph{\bibinfo{journal}{Phys. Rev. Lett.}}
  \textbf{\bibinfo{volume}{108}}, \bibinfo{pages}{145303}
  (\bibinfo{year}{2012}).

\bibitem{thuneberg_texture}
\bibinfo{author}{Thuneberg, E.~V.}
\newblock \bibinfo{title}{Hydrostatic theory of superfluid
  {$^{3}\text{He-B}$}}.
\newblock \emph{\bibinfo{journal}{J. Low Temp. Phys.}}
  \textbf{\bibinfo{volume}{122}}, \bibinfo{pages}{657--682}
  (\bibinfo{year}{2001}).

\bibitem{PhysRevB.97.014518}
\bibinfo{author}{Autti, S.}, \bibinfo{author}{Heikkinen, P.~J.},
  \bibinfo{author}{Volovik, G.~E.}, \bibinfo{author}{Zavjalov, V.~V.} \&
  \bibinfo{author}{Eltsov, V.~B.}
\newblock \bibinfo{title}{Propagation of self-localized {$Q$-ball} solitons in
  the $^{3}\mathrm{He}$ universe}.
\newblock \emph{\bibinfo{journal}{Phys. Rev. B}} \textbf{\bibinfo{volume}{97}},
  \bibinfo{pages}{014518} (\bibinfo{year}{2018}).

\bibitem{PhysRevLett.14.57}
\bibinfo{author}{Penrose, R.}
\newblock \bibinfo{title}{Gravitational collapse and space-time singularities}.
\newblock \emph{\bibinfo{journal}{Phys. Rev. Lett.}}
  \textbf{\bibinfo{volume}{14}}, \bibinfo{pages}{57--59}
  (\bibinfo{year}{1965}).

\bibitem{Sato_2011}
\bibinfo{author}{Sato, Y.} \& \bibinfo{author}{Packard, R.~E.}
\newblock \bibinfo{title}{Superfluid helium quantum interference devices:
  physics and applications}.
\newblock \emph{\bibinfo{journal}{Reports on Progress in Physics}}
  \textbf{\bibinfo{volume}{75}}, \bibinfo{pages}{016401}
  (\bibinfo{year}{2011}).

\bibitem{alex2019josephson}
\bibinfo{author}{Kreil, A. J.~E.} \emph{et~al.}
\newblock \bibinfo{title}{Josephson oscillations in a room-temperature
  {Bose-Einstein} magnon condensate}.
\newblock \emph{\bibinfo{journal}{arXiv:1911.07802}}  (\bibinfo{year}{2019}).

\bibitem{Heikkinen2014}
\bibinfo{author}{Heikkinen, P.~J.}, \bibinfo{author}{Autti, S.},
  \bibinfo{author}{Eltsov, V.~B.}, \bibinfo{author}{Haley, R.~P.} \&
  \bibinfo{author}{Zavjalov, V.~V.}
\newblock \bibinfo{title}{Microkelvin thermometry with $\text{Bose-Einstein}$
  condensates of magnons and applications to studies of the $\text{AB}$
  interface in superfluid $^{3}\text{He}$}.
\newblock \emph{\bibinfo{journal}{J. Low Temp. Phys.}}
  \textbf{\bibinfo{volume}{175}}, \bibinfo{pages}{681--705}
  (\bibinfo{year}{2014}).

\bibitem{2007_forks}
\bibinfo{author}{Blaauwgeers, R.} \emph{et~al.}
\newblock \bibinfo{title}{Quartz tuning fork: Thermometer, pressure- and
  viscometer for helium liquids}.
\newblock \emph{\bibinfo{journal}{J. Low Temp. Phys.}}
  \textbf{\bibinfo{volume}{146}}, \bibinfo{pages}{537--562}
  (\bibinfo{year}{2007}).

\bibitem{2008_forks}
\bibinfo{author}{Bla\v{z}kov\'{a}, M.} \emph{et~al.}
\newblock \bibinfo{title}{Vibrating quartz fork: A tool for cryogenic helium
  research}.
\newblock \emph{\bibinfo{journal}{J. Low Temp. Phys.}}
  \textbf{\bibinfo{volume}{150}}, \bibinfo{pages}{525--535}
  (\bibinfo{year}{2008}).

\bibitem{pjheikki_thesis}
\bibinfo{author}{Heikkinen, P.~J.}
\newblock \emph{\bibinfo{title}{Magnon {B}ose-{E}instein condensate as a probe
  of topological superfluid}}.
\newblock Ph.D. thesis, \bibinfo{school}{Aalto University School of Science}
  (\bibinfo{year}{2016}).
\newblock
  \bibinfo{note}{\url{https://aaltodoc.aalto.fi/handle/123456789/20580}}.

\bibitem{sautti_thesis}
\bibinfo{author}{Autti, S.}
\newblock \emph{\bibinfo{title}{Higgs bosons, half-quantum vortices, and
  {Q}-balls: an expedition in the $^3${H}e universe}}.
\newblock Ph.D. thesis, \bibinfo{school}{Aalto University School of Science}
  (\bibinfo{year}{2017}).
\newblock
  \bibinfo{note}{\url{https://aaltodoc.aalto.fi/handle/123456789/26282}}.

\bibitem{kopu_texture}
\bibinfo{author}{Kopu, J.}
\newblock \bibinfo{title}{Numerically calculated $\text{NMR}$ response from
  different vortex distributions in superfluid {$^{3}\text{He-B}$}}.
\newblock \emph{\bibinfo{journal}{J. Low Temp. Phys.}}
  \textbf{\bibinfo{volume}{146}}, \bibinfo{pages}{47--58}
  (\bibinfo{year}{2007}).

\end{thebibliography}
\bibliographystyle{naturemag}

\end{document}